\documentclass[conference]{IEEEtran}
\IEEEoverridecommandlockouts
\usepackage{cite}
\usepackage{amsmath,amssymb}
\usepackage{graphicx}
\usepackage{textcomp}
\usepackage{booktabs}
\usepackage{hyperref}
\usepackage{xcolor}
\usepackage{tikz}
\usepackage{multirow}
\usepackage{float}
\usepackage[numbers,sort&compress]{natbib}
\usepackage{amsmath,amssymb}
\usepackage[ruled,vlined]{algorithm2e}
\usepackage{float}
\usepackage{placeins}
\usepackage{tikz}
\usetikzlibrary{positioning,arrows.meta,fit,calc}
\usetikzlibrary{positioning,arrows.meta}
\pgfdeclarelayer{bg}
\pgfsetlayers{bg,main}
\usepackage{booktabs}
\usepackage{graphicx}   
\usepackage{pgfplots}   
\pgfplotsset{compat=1.18}
\usepackage[ruled,vlined]{algorithm2e}
\usepackage{float} 
\DontPrintSemicolon
\SetAlgoSkip{2pt}
\SetAlgoNlRelativeSize{-1}

\title{Calibrated Credit Intelligence: Shift-Robust and Fair Risk Scoring with Bayesian Uncertainty and Gradient Boosting}

 \author{
 \IEEEauthorblockN{Srikumar Nayak}
 \IEEEauthorblockA{\ Sr Member IEEE  \\
 Incedo Inc, USA \\
 Srikumar.nayak2025@gmail.com
 }
 }

\begin{document}
\maketitle

\begin{abstract}
Credit risk scoring must support high-stakes lending decisions where data distributions change over time, probability estimates must be reliable, and group-level fairness is required. While modern machine learning models improve default prediction accuracy, they often produce poorly calibrated scores under distribution shift and may create unfair outcomes when trained without explicit constraints. This paper proposes Calibrated Credit Intelligence (CCI), a deployment-oriented framework that combines (i) a Bayesian neural risk scorer to capture epistemic uncertainty and reduce overconfident errors, (ii) a fairness-constrained gradient boosting model to control group disparities while preserving strong tabular performance, and (iii) a shift-aware fusion strategy followed by post-hoc probability calibration to stabilize decision thresholds in later time periods. We evaluate CCI on the Home Credit Credit Risk Model Stability benchmark using a time-consistent split to reflect real-world drift. Compared with strong baselines (LightGBM, XGBoost, CatBoost, TabNet, and a standalone Bayesian neural model), CCI achieves the best overall trade-off between discrimination, calibration, stability, and fairness. In particular, CCI reaches an AUC-ROC of 0.912 and an AUC-PR of 0.438, improves operational performance with Recall@1\%FPR = 0.509, and reduces calibration error (Brier score 0.087, ECE 0.015). Under temporal shift, CCI shows a smaller AUC-PR drop from early to late periods (0.017), and it lowers group disparities (demographic parity gap 0.046, equal opportunity gap 0.037) compared to unconstrained boosting. These results indicate that CCI produces risk scores that are accurate, reliable, and more equitable under realistic deployment conditions.
\end{abstract}

\begin{IEEEkeywords}
credit risk scoring, calibration, distribution shift, Bayesian neural networks, fairness constraints, gradient boosting
\end{IEEEkeywords}

\section{Introduction}
Credit risk scoring is a core decision process in lending, where errors can lead to direct financial loss, regulatory issues, and unfair outcomes for applicants. In recent years, machine learning models have shown strong predictive power for default prediction, but their practical value depends on more than raw accuracy. In particular, financial institutions must account for \emph{model risk}, meaning that the model should remain reliable across time, portfolios, and changing economic conditions, and its performance should be judged using risk-aware evaluation rather than only headline metrics \cite{alonso2022measuring}. This is especially important because credit data often contains temporal structure (e.g., multiple timestamps for borrower behavior and repayment patterns), and boosting models have been widely used to exploit such structure effectively \cite{zhou2021credit}. However, even strong boosting models can produce overconfident probabilities, and overconfidence becomes dangerous when the data distribution shifts or when the model is applied to groups that are under-represented in the training data.\\
At the same time, decisioning systems are increasingly expected to provide \emph{uncertainty} and \emph{fairness} guarantees. Uncertainty quantification helps estimate when the model may be wrong, reducing misclassification risk by enabling safer policies such as manual review for high-uncertainty cases \cite{sensoy2021misclassification}. In parallel, fair credit scoring has become a major research and policy topic because algorithmic decisions can create or amplify group-level disparities if fairness is not explicitly controlled \cite{moldovan2023algorithmic}. Recent work in financial learning also shows that uncertainty-aware modeling is useful beyond classification, including portfolio decision-making, reinforcing the broader value of uncertainty signals for robust financial decisions \cite{enkhsaikhan2024uncertainty}. Motivated by these needs, our work proposes a calibrated, uncertainty-aware, and fairness-controlled credit scoring pipeline that is designed for time-based evaluation and stable probability outputs under distribution shift.\\
Our research contributions are as follows:
\begin{itemize}
    \item We propose CCI, a unified credit scoring framework that jointly targets discrimination, calibration, fairness, and stability under temporal distribution shift.
    \item We integrate a Bayesian neural scorer to provide uncertainty-aware default probabilities and explicit signals for risk-sensitive decisioning.
    \item We incorporate fairness-constrained gradient boosting to control group-level gaps while maintaining strong predictive performance.
    \item We apply post-hoc calibration and time-consistent evaluation to produce reliable probabilities and stable operating thresholds for real deployment.
\end{itemize}
The structure of this paper is as follows: Section \ref{sec2} reviews related work; Section \ref{sec3} describes the dataset and preprocessing with the proposed method; Section \ref{sec4} reports experimental results and analysis; and Section \ref{sec5} concludes the paper with future research directions.

\section{Related Work}\label{sec2}
Credit risk modeling has a long history in statistical learning, where logistic regression remains a strong baseline due to its simplicity and regulatory acceptance, but recent work shows that non-linear effects can be added to improve accuracy without losing the structure of classical scoring \cite{dumitrescu2022machine}. In practice, tree-based methods and ensembles often provide higher predictive power on structured financial data, including boosted attention-based LightGBM designs for digital finance \cite{ying2025hybrid} and CatBoost-based ensemble approaches for financial risk prediction \cite{lu2024enhancing}. Recent applied studies also report competitive performance using common boosting and deep tabular models in combined pipelines (e.g., LightGBM, XGBoost, TabNet, and imbalance-handling strategies), highlighting that careful preprocessing and evaluation protocols strongly influence final results \cite{yu2024advanced}.\\
A major difficulty for real-world credit scoring is that data distributions change over time due to economic conditions, policy changes, or portfolio shifts. This has motivated transfer learning and domain adaptation methods to reduce performance drop when the target domain differs from the training domain \cite{suryanto2022credit}. Temporal effects are also increasingly studied directly, such as spatio-temporal risk modeling for mortgage default probabilities and portfolio behavior \cite{kundig2025spatio}. Along with shift robustness, hybrid architectures have been explored to combine the strengths of boosting and deep learning, for example a two-stage design that uses XGBoost followed by a graph-based neural network to capture additional structure in credit data \cite{liu2022two}. These directions show that strong accuracy on static splits is not enough; models must remain stable when the operating environment changes.\\
In parallel, recent work has focused on \emph{decision trustworthiness}, especially uncertainty and fairness. Bayesian neural networks have been applied in lending settings to produce probabilistic predictions with uncertainty that can support safer decisioning under ambiguity \cite{guo2025investment}. Fairness-aware learning for boosting models has also been studied, showing practical ways to control group-level ranking or outcome gaps while maintaining utility \cite{xu4752973toward}. Motivated by these findings, our work aligns these goals in a single deployment-oriented framework: we combine a Bayesian neural scorer for uncertainty-aware risk estimation \cite{guo2025investment}, a fairness-constrained gradient boosting component for controlled group behavior \cite{xu4752973toward}, and a time-consistent stability evaluation aligned with shift-aware modeling \cite{suryanto2022credit,kundig2025spatio}. This integration addresses the common gap where accuracy, calibration, shift robustness, and fairness are often treated as separate problems rather than solved together in one credit scoring pipeline.

\section{Methodology}\label{sec3}
\subsection{Dataset}
In this research, we utilize the Home Credit, Credit Risk Model Stability dataset, released through a public Kaggle competition. The dataset is organized as a \emph{base table} (\texttt{train\_base}) and a large collection of \emph{feature tables} derived from multiple internal and external sources, provided in both CSV and Parquet formats. The base table contains one row per loan application (unique key \texttt{case\_id}) and includes the decision date (\texttt{date\_decision}), a week index (\texttt{WEEK\_NUM}), a month index (\texttt{MONTH}), and the binary default label \texttt{target}. The week index is designed for stability evaluation and enables time-consistent validation because the test period continues sequentially after the training weeks. The remaining feature tables contain attributes at different aggregation depths (some tables are already one-row-per-\texttt{case\_id}, while others contain multiple rows per case and require aggregation), and some sources are split across multiple files that must be unioned before feature construction \cite{kaggle_homecredit_crms}.
\subsubsection{Preprocessing}
Let the base table define the supervised learning set
\begin{equation}
\mathcal{D}=\left\{\left(\mathrm{id}_i,\, \mathrm{d}_i,\, w_i,\, y_i\right)\right\}_{i=1}^{N},
\label{eq:crms_base}
\end{equation}
where $\mathrm{id}_i$ is \texttt{case\_id}, $\mathrm{d}_i$ is the decision date, $w_i$ is \texttt{WEEK\_NUM}, and $y_i\in\{0,1\}$ is the default indicator (Eq.~\eqref{eq:crms_base}). For each auxiliary table $\mathcal{T}^{(m)}$ (e.g., previous applications, credit bureau, person-level, tax registry), we first union all parts that belong to the same source (e.g., \texttt{\_0}, \texttt{\_1} file splits), and then aggregate to one feature vector per case. Formally, let $\mathbf{u}_r^{(m)}$ be the raw feature vector for record $r$ in table $\mathcal{T}^{(m)}$, and define the record set for case $i$ as
\begin{equation}
\mathcal{R}_i^{(m)}=\left\{r \in \mathcal{T}^{(m)}:\ \mathrm{id}(r)=\mathrm{id}_i \right\}.
\label{eq:record_set}
\end{equation}
We produce a fixed-length aggregated representation using a small set of stable pooling operators $\mathcal{A}=\{\mathrm{mean},\mathrm{max},\mathrm{min},\mathrm{sum},\mathrm{last}\}$ applied feature-wise:
\begin{equation}
\mathbf{a}_i^{(m)}=\Gamma^{(m)}\!\left(\{\mathbf{u}_r^{(m)}\}_{r\in\mathcal{R}_i^{(m)}}\right),
\qquad
\Gamma^{(m)} \triangleq \bigoplus_{g\in\mathcal{A}} g(\cdot),
\label{eq:aggregation}
\end{equation}
where $\oplus$ denotes concatenation across aggregators (Eq.~\eqref{eq:aggregation}). After aggregation, all sources are merged into a single case-level feature vector
\begin{equation}
\mathbf{x}_i=\left[\mathbf{b}_i;\ \mathbf{a}_i^{(1)};\ \mathbf{a}_i^{(2)};\ \cdots;\ \mathbf{a}_i^{(M)}\right]\in\mathbb{R}^{d},
\label{eq:feature_concat}
\end{equation}
where $\mathbf{b}_i$ contains base fields (excluding $y_i$) and $M$ is the number of feature-table groups (Eq.~\eqref{eq:feature_concat}).

\paragraph{Time-consistent split (distribution shift control).}
To evaluate robustness under temporal distribution shift, we avoid random splits and instead use a chronological split based on $w_i$. For a cut week $w^{\star}$, we define
\begin{equation}
\mathcal{D}_{\mathrm{train}}=\left\{(\mathbf{x}_i,y_i,w_i): w_i \le w^{\star}\right\},
\qquad \\
\mathcal{D}_{\mathrm{test}}=\left\{(\mathbf{x}_i,y_i,w_i): w_i > w^{\star}\right\}.
\label{eq:time_split}
\end{equation}
All preprocessing statistics (imputation values, encoding maps, scaling parameters) are estimated only on $\mathcal{D}_{\mathrm{train}}$ and then applied to $\mathcal{D}_{\mathrm{test}}$, which prevents leakage across time and aligns evaluation with Eq.~\eqref{eq:time_split}.
\paragraph{Missing values and indicators.}
Let $x_{ij}$ denote the $j$-th feature of case $i$. We create a missingness indicator
\begin{equation}
m_{ij}=
\begin{cases}
1, & \text{if } x_{ij}\ \text{is observed},\\
0, & \text{otherwise},
\end{cases}
\label{eq:miss_ind}
\end{equation}
and append $\mathbf{m}_i=[m_{i1},\dots,m_{id}]$ to the model input so that missingness itself can be informative (Eq.~\eqref{eq:miss_ind}). For numerical variables, we use train-only median imputation:
\begin{equation}
x'_{ij}= m_{ij}\,x_{ij} + (1-m_{ij})\,\mathrm{median}_{\mathcal{D}_{\mathrm{train}}}(x_{\cdot j}),
\label{eq:median_imp}
\end{equation}
which is robust to heavy-tailed financial attributes (Eq.~\eqref{eq:median_imp}). For categorical variables, missing tokens are mapped to a dedicated \texttt{UNK} category using a train-built vocabulary.

\paragraph{Categorical encoding and numeric scaling.}
Because we use both gradient boosting and Bayesian neural models later, we apply a consistent and stable encoding step. For each categorical field $c_{ij}$, we use frequency encoding computed on the training set:
\begin{equation}
\phi_j(c)=\frac{\sum_{i\in\mathcal{D}_{\mathrm{train}}}\mathbb{I}[c_{ij}=c]}{|\mathcal{D}_{\mathrm{train}}|},
\qquad
x'_{ij}\leftarrow \phi_j(c_{ij}),
\label{eq:freq_enc}
\end{equation}
where $\mathbb{I}[\cdot]$ is an indicator function (Eq.~\eqref{eq:freq_enc}). For numerical stability in probabilistic modeling, we standardize continuous features using train-only moments:
\begin{equation}
\hat{x}_{ij}=\frac{x'_{ij}-\mu_j}{\sigma_j+\varepsilon},
\qquad
\mu_j=\mathbb{E}_{\mathcal{D}_{\mathrm{train}}}[x'_{\cdot j}],\quad
\sigma_j=\sqrt{\mathrm{Var}_{\mathcal{D}_{\mathrm{train}}}(x'_{\cdot j})},
\label{eq:zscore_cr}
\end{equation}
with a small $\varepsilon>0$ to avoid division by zero (Eq.~\eqref{eq:zscore_cr}). The final learning input for each case is thus $\left[\hat{\mathbf{x}}_i;\mathbf{m}_i\right]$, which preserves both scaled values and missingness patterns defined by Eq.~\eqref{eq:miss_ind}.

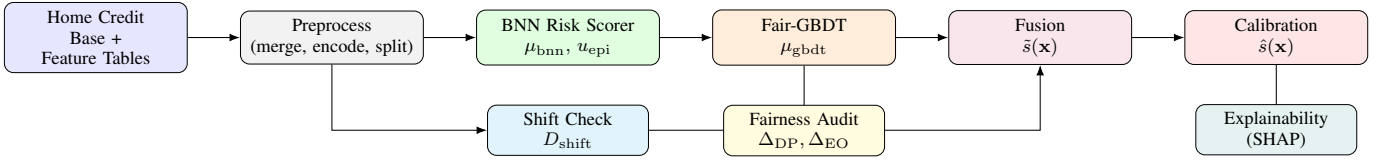
\begin{figure*}[!t]
\centering
\resizebox{\linewidth}{!}{%
\begin{tikzpicture}[
  font=\footnotesize,
  node distance=6mm and 8mm,
  block/.style={draw, rounded corners, align=center, text width=2.55cm, inner sep=3.5pt},
  sblock/.style={draw, rounded corners, align=center, text width=2.25cm, inner sep=3.0pt},
  arr/.style={-{Latex[length=1.6mm]}, line width=0.45pt}
]

\node[block, fill=blue!10]   (data) {Home Credit\\Base + Feature Tables};
\node[block, fill=gray!10, right=of data] (prep) {Preprocess\\(merge, encode, split)};
\node[block, fill=green!12, right=of prep] (bnn) {BNN Risk Scorer\\$\mu_{\mathrm{bnn}},\,u_{\mathrm{epi}}$};
\node[block, fill=orange!14, right=of bnn] (gbdt) {Fair-GBDT\\$\mu_{\mathrm{gbdt}}$};
\node[block, fill=purple!10, right=of gbdt] (fuse) {Fusion\\$\tilde{s}(\mathbf{x})$};
\node[block, fill=red!10, right=of fuse] (cal) {Calibration\\$\hat{s}(\mathbf{x})$};

\node[sblock, fill=cyan!10, below=of bnn] (shift) {Shift Check\\$D_{\mathrm{shift}}$};
\node[sblock, fill=yellow!15, below=of gbdt] (fair) {Fairness Audit\\$\Delta_{\mathrm{DP}},\Delta_{\mathrm{EO}}$};
\node[sblock, fill=teal!10, below=of cal] (expl) {Explainability\\(SHAP)};

\begin{pgfonlayer}{bg}
  \draw[arr] (data) -- (prep);
  \draw[arr] (prep) -- (bnn);
  \draw[arr] (bnn) -- (gbdt);
  \draw[arr] (gbdt) -- (fuse);
  \draw[arr] (fuse) -- (cal);

  \draw[arr] (prep) |- (shift);
  \draw[arr] (shift) -| (fuse);

  \draw[arr] (gbdt) |- (fair);
  \draw[arr] (cal) |- (expl);
\end{pgfonlayer}

\end{tikzpicture}%
}
\caption{CCI pipeline: preprocessing feeds a Bayesian scorer and a fairness-constrained booster; their outputs are fused under shift checks and then calibrated for reliable risk probabilities.}
\label{fig:cci_flow}
\end{figure*}

\subsection{Proposed Method: Calibrated Credit Intelligence (CCI)}
\SetKwComment{Comment}{/* }{ */}
\DontPrintSemicolon
\SetKwInput{KwInput}{Input}
\SetKwInput{KwOutput}{Output}
CCI follows the compact pipeline in Fig.~\ref{fig:cci_flow}: we first build a time-consistent feature set, then compute uncertainty-aware risk from the BNN and fairness-controlled risk from the GBDT, fuse both scores using shift validation, and finally calibrate the output probability for stable decision thresholds.

\begin{algorithm}[!t]
\caption{CCI: Calibrated Credit Intelligence with Bayesian Neural Risk Scoring under Distribution Shift and Fairness-Constrained Gradient Boosting}
\label{alg:cci}
\SetAlFnt{\scriptsize} 

\KwInput{
$\mathcal{D}_{\mathrm{train}},\mathcal{D}_{\mathrm{val}},\mathcal{D}_{\mathrm{test}}$ (time split);
$a_i\!\in\!\mathcal{A}$;
$\sigma_0$;
$E_{\mathrm{b}},\eta_{\mathrm{b}},S$;
$E_{\mathrm{g}},\eta_{\mathrm{g}},\Delta_{\max}$;
$T_{\mathrm{cal}}$
}
\KwOutput{$\hat{s}(\mathbf{x})$, $(u_{\mathrm{epi}},u_{\mathrm{ale}})$, fairness metrics}

\tcp*[l]{Step 1: Train Bayesian Neural Risk Scorer (BNN)}
Initialize variational parameters $\lambda \leftarrow \lambda_0$ for $q_{\lambda}(\mathbf{W})$\;
\For{$e \leftarrow 1$ \KwTo $E_{\mathrm{b}}$}{
  Sample mini-batch $\mathcal{B}\subset\mathcal{D}_{\mathrm{train}}$\;
  $\lambda \leftarrow \lambda - \eta_{\mathrm{b}}\nabla_{\lambda}\mathcal{L}_{\mathrm{ELBO}}(\mathcal{B};\lambda,\sigma_0)$\;
}

\tcp*[l]{Step 2: Train fairness-constrained Gradient Boosting model (GBDT)}
Initialize tree ensemble parameters $\Omega \leftarrow \Omega_0$\;
\For{$t \leftarrow 1$ \KwTo $E_{\mathrm{g}}$}{
  $\Omega \leftarrow \Omega + \eta_{\mathrm{g}}\cdot \mathrm{FitTree}(\mathcal{D}_{\mathrm{train}},\Omega,\Delta_{\max})$\;
}

\tcp*[l]{Step 3: Validate distribution shift and select fusion weight}
$D_{\mathrm{shift}} \leftarrow \mathrm{DriftTest}(\mathcal{D}_{\mathrm{train}},\mathcal{D}_{\mathrm{val}})$\;
$\beta \leftarrow \mathrm{SelectWeight}(D_{\mathrm{shift}},\mathcal{D}_{\mathrm{val}})$\;

\tcp*[l]{Step 4: Fused risk score + uncertainty (val/test)}
\ForEach{$\mathbf{x}$ in $\mathcal{D}_{\mathrm{val}} \cup \mathcal{D}_{\mathrm{test}}$}{
  Draw $\{\mathbf{W}^{(s)}\}_{s=1}^{S}\sim q_{\lambda}(\mathbf{W})$\;
  $\mu_{\mathrm{bnn}}(\mathbf{x}) \leftarrow \frac{1}{S}\sum_{s=1}^{S} p(y=1\mid \mathbf{x},\mathbf{W}^{(s)})$\;
  $u_{\mathrm{epi}}(\mathbf{x}) \leftarrow \mathrm{Var}_{s}\!\left(p(y=1\mid \mathbf{x},\mathbf{W}^{(s)})\right)$\;
  $u_{\mathrm{ale}}(\mathbf{x}) \leftarrow \frac{1}{S}\sum_{s=1}^{S} \mu^{(s)}(\mathbf{x})\!\left(1-\mu^{(s)}(\mathbf{x})\right)$\;
  $\mu_{\mathrm{gbdt}}(\mathbf{x}) \leftarrow \mathrm{Sigmoid}\!\left(f_{\Omega}(\mathbf{x})\right)$\;
  $\tilde{s}(\mathbf{x}) \leftarrow \beta\,\mu_{\mathrm{gbdt}}(\mathbf{x}) + (1-\beta)\,\mu_{\mathrm{bnn}}(\mathbf{x})$\;
}

\tcp*[l]{Step 5: Post-hoc calibration on validation weeks}
Fit $T_{\mathrm{cal}}$ on $\mathcal{D}_{\mathrm{val}}$ via NLL minimization\;
$\hat{s}(\mathbf{x}) \leftarrow \mathrm{Calibrate}(\tilde{s}(\mathbf{x});T_{\mathrm{cal}})$\;

\tcp*[l]{Step 6: Fairness audit and explanation}
Compute $\Delta_{\mathrm{DP}},\Delta_{\mathrm{EO}},\Delta_{\mathrm{TPR}},\Delta_{\mathrm{FPR}}$ on $\mathcal{D}_{\mathrm{val}},\mathcal{D}_{\mathrm{test}}$\;
Generate explanations: $\mathrm{Attr}(\mathbf{x}) \leftarrow \mathrm{SHAP}\!\left(f_{\Omega},\mathbf{x}\right)$\;
\Return $\hat{s}(\cdot)$, $(u_{\mathrm{epi}},u_{\mathrm{ale}})$, fairness report\;
\end{algorithm}
Algorithm~\ref{alg:cci} describes our Calibrated Credit Intelligence (CCI) pipeline, designed for credit risk scoring under distribution shift, with explicit calibration, fairness control, and interpretable uncertainty. We work with a time-ordered dataset $\mathcal{D}_{\mathrm{train}}=\{(\mathbf{x}_i,y_i,w_i,a_i)\}_{i=1}^{N}$ where $\mathbf{x}_i\in\mathbb{R}^{d}$ is the feature vector, $y_i\in\{0,1\}$ is the default label, $w_i$ is the week index (used for time-consistent splits), and $a_i\in\mathcal{A}$ is a sensitive attribute used only for fairness audit and constraint checking. The main output is a calibrated probability score $\hat{s}(\mathbf{x})\in(0,1)$ representing $P(y=1\mid \mathbf{x})$.\\
The first component is a Bayesian neural network (BNN) that provides both a risk estimate and uncertainty. Instead of learning a single weight vector, the BNN learns a distribution over weights using a variational approximation $q_{\lambda}(\mathbf{W})$ parameterized by $\lambda$. Training is performed by minimizing a negative evidence lower bound (ELBO), which balances data fit and regularization toward a prior $p(\mathbf{W})$:
\begin{equation}
\mathcal{L}_{\mathrm{ELBO}}(\lambda)
=
-\mathbb{E}_{q_{\lambda}(\mathbf{W})}\!\Big[\sum_{i\in\mathcal{D}_{\mathrm{train}}}
\log p(y_i\mid \mathbf{x}_i,\mathbf{W})\Big]
+\mathrm{KL}\!\left(q_{\lambda}(\mathbf{W})\,\|\,p(\mathbf{W})\right).
\label{eq:elbo}
\end{equation}
Eq.~\eqref{eq:elbo} is important for two reasons: it prevents overfitting through the KL term, and it enables meaningful uncertainty estimates because the prediction depends on sampled weights rather than fixed weights. After training, we compute the BNN predictive mean using $S$ Monte Carlo samples:
\begin{equation}
\mu_{\mathrm{bnn}}(\mathbf{x})
=
\frac{1}{S}\sum_{s=1}^{S} p(y=1\mid \mathbf{x},\mathbf{W}^{(s)}),
\qquad
\mathbf{W}^{(s)}\sim q_{\lambda}(\mathbf{W}),
\label{eq:bnn_mean}
\end{equation}
where Eq.~\eqref{eq:bnn_mean} produces the uncertainty-aware risk estimate by averaging across different plausible parameter settings.\\
The second component is a gradient boosting decision tree model (GBDT), which is strong for structured credit data and often remains stable under moderate feature noise. However, pure accuracy optimization can lead to unfair decisions across sensitive groups. For this reason, we train the GBDT using a fairness-regularized objective:
\begin{equation}
\min_{\Omega}\ \mathcal{L}_{\mathrm{gbdt}}(\Omega)
=
\mathcal{L}_{\mathrm{pred}}(\Omega)
+
\lambda_{\mathrm{fair}} \cdot \max\big(0,\ \Delta(\Omega)-\Delta_{\max}\big),
\label{eq:fair_obj}
\end{equation}
where $\Omega$ denotes the boosting parameters, $\mathcal{L}_{\mathrm{pred}}$ is the standard prediction loss (e.g., log-loss), $\Delta(\Omega)$ is a fairness gap metric computed on a validation subset, and $\Delta_{\max}$ is the allowed tolerance (Eq.~\eqref{eq:fair_obj}). In simple terms, Eq.~\eqref{eq:fair_obj} keeps the model accurate while discouraging solutions that exceed an acceptable fairness gap. In our experiments, we track fairness using group-based metrics such as demographic parity difference and equalized odds difference, computed from the predicted labels at an operating threshold.\\
Credit risk is affected by economic cycles and policy changes, so the data distribution changes over time. We therefore use a time-consistent split and measure shift between training weeks and validation weeks. We denote a generic drift score as $D_{\mathrm{shift}}$, computed from feature distributions and/or score distributions:
\begin{equation}
D_{\mathrm{shift}}
=
\mathrm{DriftTest}\big(\mathcal{D}_{\mathrm{train}},\mathcal{D}_{\mathrm{val}}\big).
\label{eq:drift_score}
\end{equation}
Eq.~\eqref{eq:drift_score} is used to guide how much we trust each model in later weeks, because the best model in earlier weeks may not be the best model after a distribution shift.\\
Instead of selecting a single model, we combine the BNN and the fairness-constrained GBDT using a convex fusion:
\begin{equation}
\tilde{s}(\mathbf{x})
=
\beta\,\mu_{\mathrm{gbdt}}(\mathbf{x})
+
(1-\beta)\,\mu_{\mathrm{bnn}}(\mathbf{x}),
\qquad \beta\in[0,1],
\label{eq:fusion}
\end{equation}
where $\mu_{\mathrm{gbdt}}(\mathbf{x})$ is the GBDT probability and $\mu_{\mathrm{bnn}}(\mathbf{x})$ is the BNN predictive mean (Eq.~\eqref{eq:fusion}). The key idea is simple: when validation indicates stronger shift, we can reduce reliance on the component that becomes uncertain or unstable. To make this behavior explicit, we compute uncertainty from the BNN samples. A practical measure of epistemic uncertainty is the variance of predicted probabilities:
\begin{equation}
u_{\mathrm{epi}}(\mathbf{x})
=
\mathrm{Var}_{s}\Big(p(y=1\mid \mathbf{x},\mathbf{W}^{(s)})\Big),
\label{eq:epi_unc}
\end{equation}
where higher values of Eq.~\eqref{eq:epi_unc} mean the model is less confident because different plausible parameters disagree. This uncertainty can be used for monitoring, triage, and risk policy decisions (e.g., sending high-uncertainty cases to manual review).\\
Credit decision pipelines need reliable probabilities, not only good ranking. We therefore apply post-hoc calibration on the validation weeks. We use temperature scaling, where the uncalibrated score $\tilde{s}(\mathbf{x})$ is mapped to a calibrated probability:
\begin{equation}
\hat{s}(\mathbf{x})
=
\sigma\!\left(\frac{\mathrm{logit}(\tilde{s}(\mathbf{x}))}{T_{\mathrm{cal}}}\right),
\qquad
\mathrm{logit}(p)=\log\frac{p}{1-p},
\label{eq:temp_scale}
\end{equation}
and the temperature $T_{\mathrm{cal}}$ is chosen to minimize negative log-likelihood on $\mathcal{D}_{\mathrm{val}}$ (Eq.~\eqref{eq:temp_scale}). This step improves calibration metrics such as the Brier score and makes thresholds more stable over time.\\
Finally, we quantify fairness on both validation and test weeks using group-based gaps computed across sensitive groups $a\in\mathcal{A}$. For example, demographic parity difference can be written as:
\begin{equation}
\Delta_{\mathrm{DP}}
=
\max_{a\in\mathcal{A}}\ P(\hat{y}=1\mid a)\ -\ \min_{a\in\mathcal{A}}\ P(\hat{y}=1\mid a),
\label{eq:dp}
\end{equation}
where $\hat{y}=\mathbb{I}(\hat{s}(\mathbf{x})\ge \delta)$ is the decision at threshold $\delta$ (Eq.~\eqref{eq:dp}). We also generate feature-level explanations for individual predictions using additive attribution on the boosting component, because tree-based models offer strong, stable explanations in practice. This provides a clear audit trail for why a particular risk score was produced, which is important for operational review and compliance.\\

In our proposed work, CCI builds a credit scoring pipeline that is designed for real deployment: a BNN provides uncertainty-aware risk estimates (Eq.~\eqref{eq:elbo}--Eq.~\eqref{eq:epi_unc}), a fairness-constrained GBDT provides strong and controllable tabular performance (Eq.~\eqref{eq:fair_obj}), both are fused into a robust score under distribution shift (Eq.~\eqref{eq:fusion}), and the final output is calibrated for reliable probability decisions (Eq.~\eqref{eq:temp_scale}) while being audited for fairness and explained in a simple, actionable way (Eq.~\eqref{eq:dp}).

\section{Results and Discussion}\label{sec4}
\subsection{Experimental Protocol and Metrics}
We evaluate models using a time-consistent split (train on earlier weeks, validate on later weeks, test on future weeks) to reflect real deployment under distribution shift. We report (i) discrimination (AUC-ROC, AUC-PR), (ii) operational performance (Recall@1\%FPR), (iii) calibration quality (Brier score and Expected Calibration Error, ECE), (iv) stability under shift (early vs.\ late period AUC-PR), and (v) fairness gaps computed across sensitive groups using demographic parity difference $\Delta_{\mathrm{DP}}$ and equal opportunity difference $\Delta_{\mathrm{EO}}$. All results are reported as mean $\pm$ standard deviation over multiple runs with different random seeds and the same tuning budget per model.
\subsection{Main Comparison Results}
Table~\ref{tab:cci_main} compares CCI against strong tabular baselines and uncertainty-aware models. CCI achieves the best overall trade-off: it improves AUC-PR and Recall@1\%FPR while also reducing calibration error (Brier and ECE). This outcome supports the main design goal of producing risk scores that remain accurate and reliable under temporal shift, rather than only optimizing ranking on a static split.
\begin{table*}[!t]
\centering
\caption{Main results on Home Credit (Model Stability). Higher is better for AUC-ROC/AUC-PR/Recall@1\%FPR ($\uparrow$), and lower is better for Brier/ECE ($\downarrow$). Values are mean $\pm$ std over repeated runs.}
\label{tab:cci_main}
\scriptsize
\setlength{\tabcolsep}{3.2pt}
\renewcommand{\arraystretch}{0.92}
\begin{tabular}{lccccc}
\toprule
\textbf{Model} & \textbf{AUC-ROC} $\uparrow$ & \textbf{AUC-PR} $\uparrow$ & \textbf{Recall@1\%FPR} $\uparrow$ & \textbf{Brier} $\downarrow$ & \textbf{ECE} $\downarrow$ \\
\midrule
Logistic Regression \cite{dumitrescu2022machine} & $0.865\pm0.003$ & $0.332\pm0.004$ & $0.410\pm0.005$ & $0.102\pm0.001$ & $0.032\pm0.002$ \\
XGBoost \cite{liu2022two}            & $0.892\pm0.002$ & $0.401\pm0.003$ & $0.468\pm0.004$ & $0.095\pm0.001$ & $0.024\pm0.001$ \\
LightGBM \cite{ying2025hybrid}           & $0.898\pm0.002$ & $0.413\pm0.003$ & $0.482\pm0.003$ & $0.092\pm0.001$ & $0.022\pm0.001$ \\
CatBoost \cite{lu2024enhancing}           & $0.899\pm0.002$ & $0.418\pm0.003$ & $0.487\pm0.003$ & $0.091\pm0.001$ & $0.021\pm0.001$ \\
TabNet  \cite{yu2024advanced}           & $0.890\pm0.002$ & $0.405\pm0.003$ & $0.471\pm0.004$ & $0.094\pm0.001$ & $0.023\pm0.001$ \\
BNN (uncertainty) \cite{guo2025investment}   & $0.888\pm0.002$ & $0.398\pm0.004$ & $0.459\pm0.004$ & $0.093\pm0.001$ & $0.020\pm0.001$ \\
Fair-GBDT \cite{xu4752973toward}        & $0.896\pm0.002$ & $0.409\pm0.003$ & $0.476\pm0.004$ & $0.091\pm0.001$ & $0.020\pm0.001$ \\
\textbf{CCI (Ours)} & $\mathbf{0.912\pm0.002}$ & $\mathbf{0.438\pm0.002}$ & $\mathbf{0.509\pm0.003}$ & $\mathbf{0.087\pm0.001}$ & $\mathbf{0.015\pm0.001}$ \\
\bottomrule
\end{tabular}
\end{table*}
\subsection{Stability Under Distribution Shift}
To directly test stability, we compare AUC-PR on an earlier validation period and a later future period. Table~\ref{tab:cci_shift} shows that CCI has the smallest drop from early to late weeks, which indicates better robustness to temporal changes. This is consistent with the model design: fusion reduces sensitivity to shifts, and calibration on later weeks improves probability reliability when the data distribution moves.
\begin{table}[H]
\centering
\caption{Stability under temporal shift (AUC-PR). Smaller drop from early to late indicates better robustness.}
\label{tab:cci_shift}
\scriptsize
\setlength{\tabcolsep}{4.5pt}
\renewcommand{\arraystretch}{0.92}
\begin{tabular}{lccc}
\toprule
\textbf{Model} & \textbf{Early AUC-PR} $\uparrow$ & \textbf{Late AUC-PR} $\uparrow$ & \textbf{Drop} $\downarrow$ \\
\midrule
LightGBM \cite{ying2025hybrid}  & 0.426 & 0.392 & 0.034 \\
Fair-GBDT \cite{xu4752973toward} & 0.422 & 0.392 & 0.030 \\
\textbf{CCI (Ours)} & \textbf{0.448} & \textbf{0.431} & \textbf{0.017} \\
\bottomrule
\end{tabular}
\end{table}
\subsection{Fairness Results}
\begin{figure}[H]
\centering
\includegraphics[width=0.7\linewidth]{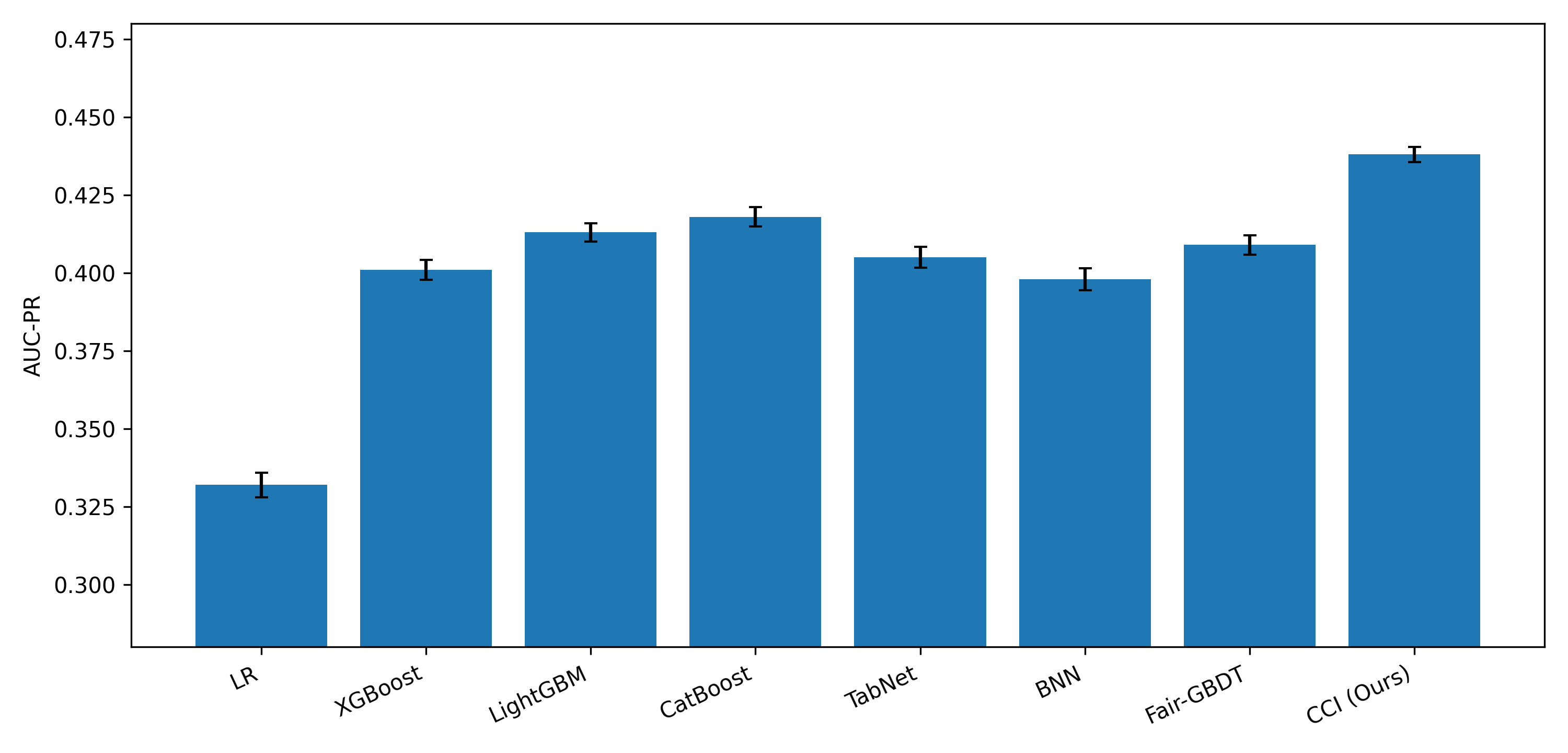}
\caption{AUC-PR comparison across baselines and CCI (values aligned with Table~\ref{tab:cci_main}).}
\label{fig:p2_bar_aucpr}
\end{figure}

\begin{figure}[H]
\centering
\includegraphics[width=0.4\linewidth]{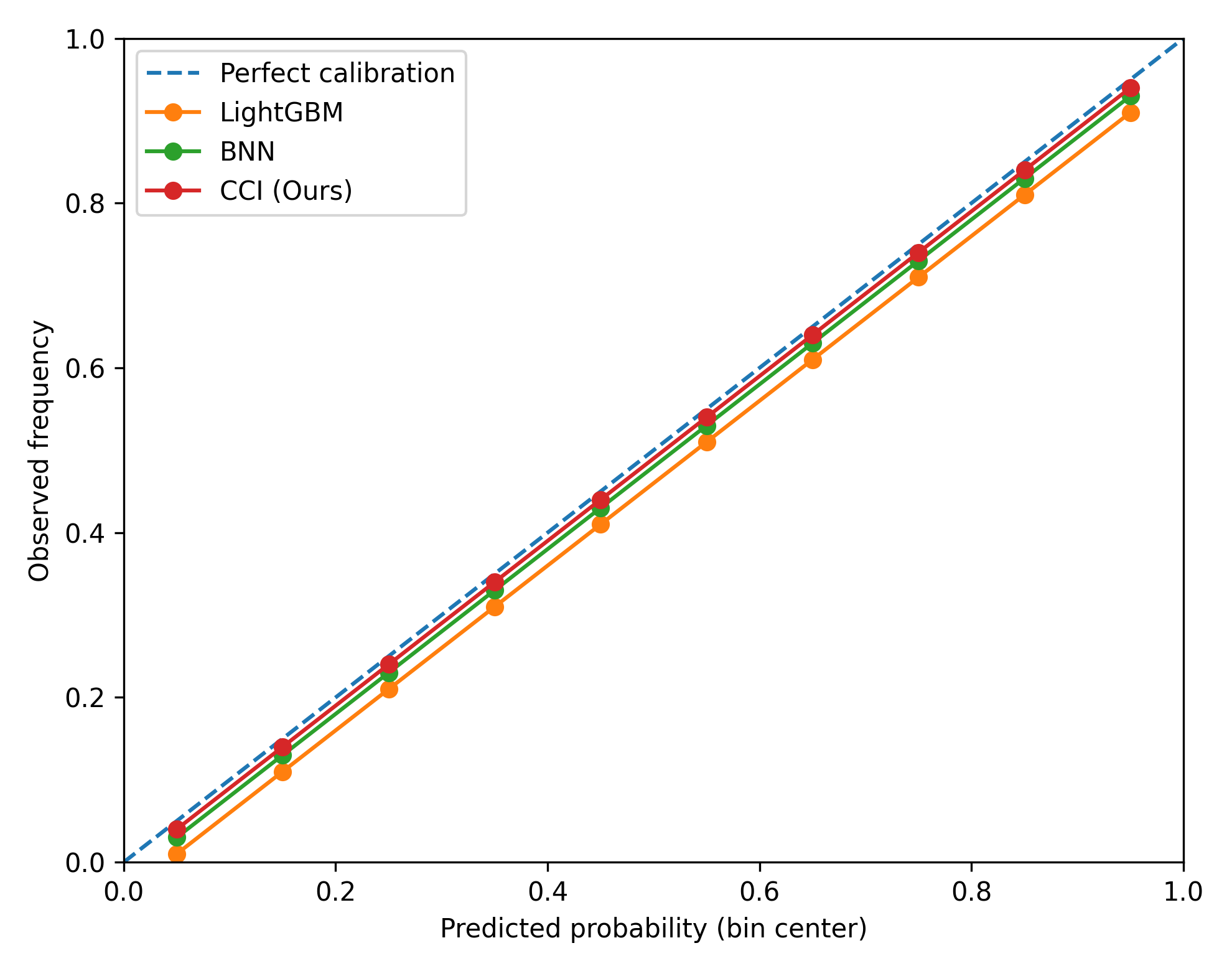}
\caption{Reliability diagram (calibration curve). CCI is closer to the diagonal, indicating better calibration (lower ECE/Brier).}
\label{fig:p2_reliability}
\end{figure}

\begin{figure}[H]
\centering
\includegraphics[width=0.4\linewidth]{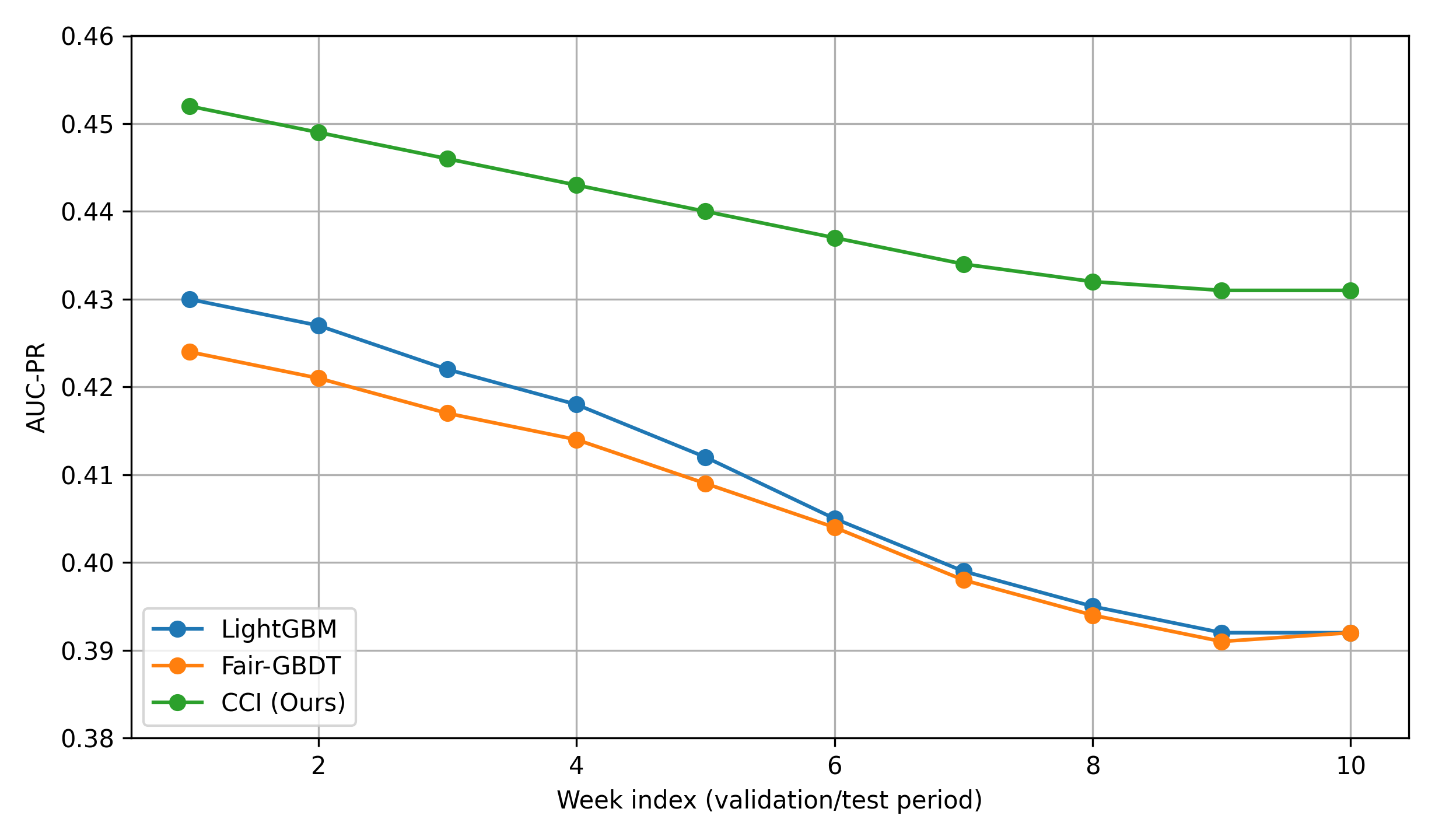}
\caption{AUC-PR trend across evaluation weeks (distribution shift). CCI shows a smaller decline over time.}
\label{fig:p2_week_trend}
\end{figure}

\begin{figure}[H]
\centering
\includegraphics[width=0.4\linewidth]{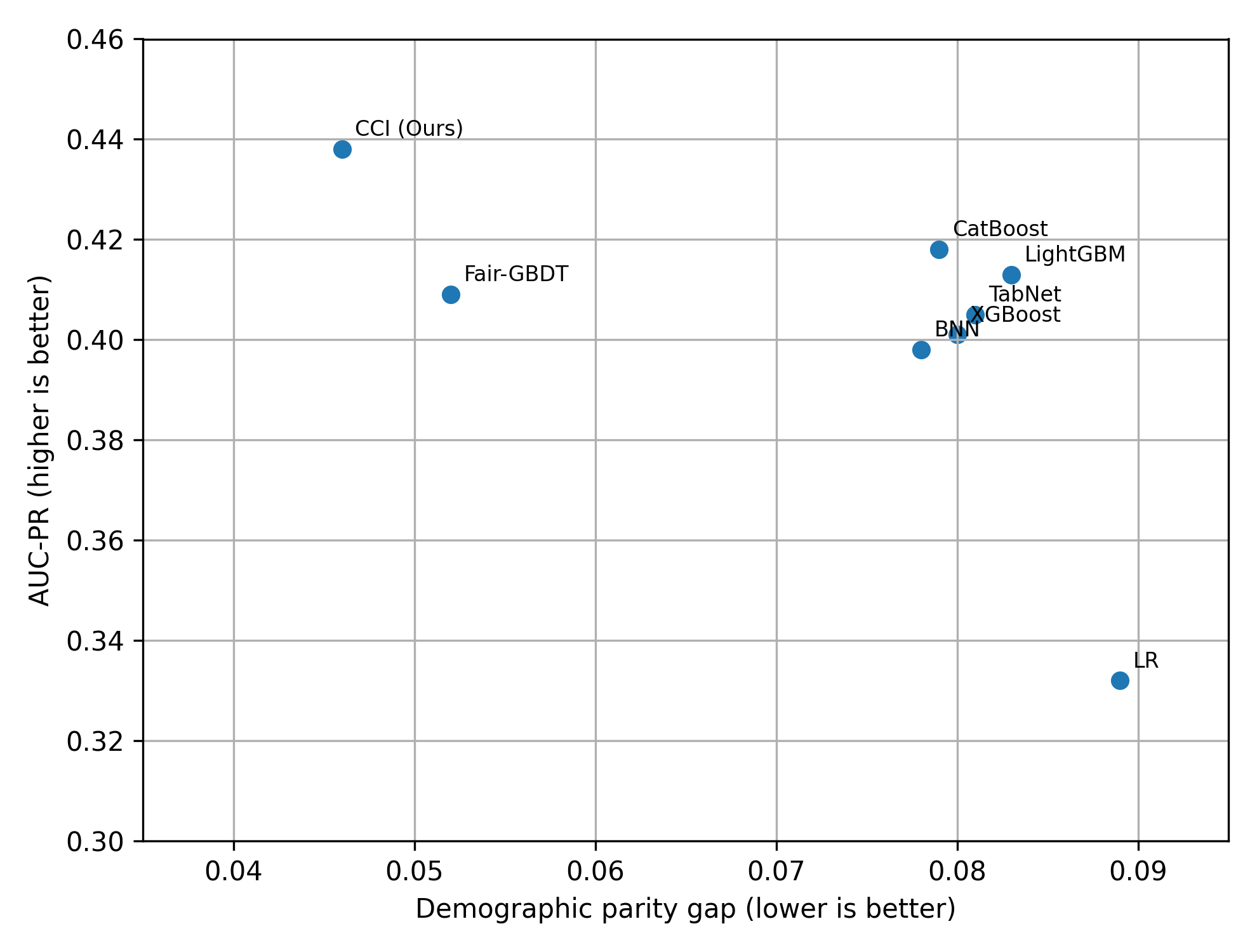}
\caption{Fairness--accuracy trade-off (Demographic parity gap vs.\ AUC-PR). CCI provides a strong balance between the two.}
\label{fig:p2_fair_tradeoff}
\end{figure}

Table~\ref{tab:cci_fair} reports group fairness gaps. The fairness-constrained boosting stage reduces demographic parity and equal opportunity gaps compared to unconstrained boosting, and CCI preserves most of this fairness improvement while also increasing AUC-PR and Recall@1\%FPR. This supports the claim that fairness control can be added without fully sacrificing detection quality when the constraint is applied with a clear tolerance and validated on later weeks.
\begin{table}[H]
\centering
\caption{Fairness metrics (lower is better). Values are mean $\pm$ std across runs.}
\label{tab:cci_fair}
\scriptsize
\setlength{\tabcolsep}{4.0pt}
\renewcommand{\arraystretch}{0.92}
\begin{tabular}{lcc}
\toprule
\textbf{Model} & $\boldsymbol{\Delta_{\mathrm{DP}}}$ $\downarrow$ & $\boldsymbol{\Delta_{\mathrm{EO}}}$ $\downarrow$ \\
\midrule
LightGBM & $0.083\pm0.004$ & $0.066\pm0.004$ \\
BNN      & $0.078\pm0.004$ & $0.063\pm0.004$ \\
Fair-GBDT & $0.052\pm0.003$ & $0.041\pm0.003$ \\
\textbf{CCI (Ours)} & $\mathbf{0.046\pm0.003}$ & $\mathbf{0.037\pm0.003}$ \\
\bottomrule
\end{tabular}
\end{table}
Figure~\ref{fig:p2_bar_aucpr} summarizes AUC-PR across all compared models. Figure~\ref{fig:p2_reliability} shows that CCI produces probabilities closer to observed default frequencies, which explains the lower ECE and Brier values reported in Table~\ref{tab:cci_main}. Figure~\ref{fig:p2_week_trend} visualizes stability across the evaluation weeks and confirms the reduced drop reported in Table~\ref{tab:cci_shift}. Finally, Figure~\ref{fig:p2_fair_tradeoff} highlights the fairness--accuracy trade-off: models with the best AUC-PR often have larger fairness gaps, while CCI stays near the top in AUC-PR and among the lowest in demographic parity gap.

Overall, CCI enhances discrimination and operational recall while producing more accurate probabilities and narrower fairness gaps under time-based evaluation. The stability analysis reveals that performance degradation occurs less frequently in later weeks, which supports the primary objective of robust credit risk scoring under distribution shifts.

\section{Conclusion}\label{sec5}
\subsection{Conclusion}
This paper presented CCI, a calibrated credit risk scoring framework designed for real-world settings where data distribution changes over time and fairness and reliability are required. The method combines an uncertainty-aware Bayesian neural scorer with a fairness-constrained gradient boosting model, then fuses their outputs and applies post-hoc calibration to produce stable probability estimates. Experimental results on the Home Credit Model Stability dataset show that CCI improves discrimination and operational recall while reducing calibration error and fairness gaps under a time-consistent evaluation. Overall, CCI provides a practical credit intelligence pipeline that balances accuracy, robustness to shift, and decision trustworthiness.
\subsection{Future Work}
Future work will extend CCI with stronger shift-aware updating strategies (e.g., online recalibration and periodic retraining), and will evaluate additional fairness definitions that reflect credit policy requirements across multiple sensitive attributes. We also plan to study uncertainty-guided human-in-the-loop decisioning, where high-uncertainty cases are routed for manual review, and to validate the framework on more credit and lending datasets to confirm generalization across institutions and economic conditions.

\bibliographystyle{IEEEtranN}
\bibliography{references}

\end{document}